\documentclass[aps,pra,twocolumn,showpacs,fleqn]{revtex4-1}

\usepackage{graphicx}

\usepackage{epsfig}
\usepackage{amsfonts}
\usepackage{amssymb}
\usepackage{mathrsfs}
\usepackage{theorem}
\usepackage{amsmath}
\usepackage{times}
\usepackage{color}
\usepackage{qcircuit}
\usepackage[T1]{fontenc}
\usepackage{ifpdf}
\ifpdf
\usepackage{epstopdf}   
\usepackage{url}
\usepackage{graphicx}
\usepackage{fancyhdr}
\usepackage{qcircuit}
\usepackage[toc,page]{appendix}
\usepackage{multirow}
\usepackage{subcaption}
\fi

\begin{document}
\title{Relative resilience to noise of standard and sequential approaches to \\ measurement-based quantum computation }
\author{C.B. Gallagher$^{1}$}
\email{cgallagher48@qub.ac.uk}
\author{A. Ferraro$^1$}
\email{a.ferraro@qub.ac.uk}
\affiliation{$^1$Centre for Theoretical Atomic, Molecular and Optical Physics,\\
School of Mathematics and Physics, Queen's University, Belfast BT7 1NN, United Kingdom}

\begin{abstract}
\normalsize
A possible alternative to the standard model of measurement-based quantum computation (MBQC) is offered by the sequential model of MBQC --- a particular class of quantum computation via ancillae.
Although these two models are equivalent under ideal conditions, their relative resilience to noise in practical conditions is not yet known. We analyze this relationship for various noise models in the ancilla preparation and in the entangling-gate implementation. The comparison of the two models is performed utilizing both the gate infidelity and the diamond distance as figures of merit. Our results show that in the majority of instances the sequential model outperforms the standard one in regard to a universal set of operations for quantum computation. Further investigation is made into the performance of sequential MBQC in experimental scenarios, thus setting benchmarks for possible cavity-QED implementations.
\end{abstract}

\maketitle

\section{Introduction}

In its standard circuital model, quantum computation is driven by a set of universal gates that act on a register of quantum systems accordingly to an adaptable pattern that depends on the algorithm to be computed \cite{nielsenchuang}. Despite the steady experimental progresses in this direction \cite{chow09,gaebler12,chow14, barends14,harty14,muhonen15,xia15,paik16}, the degree of flexibility and control required by this gate model is still demanding. Among the possible alternatives, an especially interesting one is given by \textit{measurement-based quantum computation} (MBQC) \cite{raussendorf}. The latter exploits the resources offered by a reference state composed of multiple nodes, dubbed the \textit{cluster state}; once this universal resource is built, the computation is driven by measurements, rather than gates, that involve only one node at a time and whose pattern depends on the algorithm to be implemented. The availability of such local measurements in many physical settings, and with a high degree of control, have made MBQC an attractive approach --- thus stimulating extensive efforts towards the generation of cluster states. With this aim, various schemes have been put forward, both in the context of finite \cite{RB01a,CMD+10,Miy11,WAR11} and infinite-dimensional quantum systems \cite{ZB06,M+06,A+11,HAF15}. Major experimental breakthroughs include, in the single-photon domain, the implementation of small instances of the Deutsch algorithm \cite{W+05}, and, in the continuous-variable regime, the generation of clusters comprising as many as $10^6$ time-encoded modes of travelling light \cite{furu}.

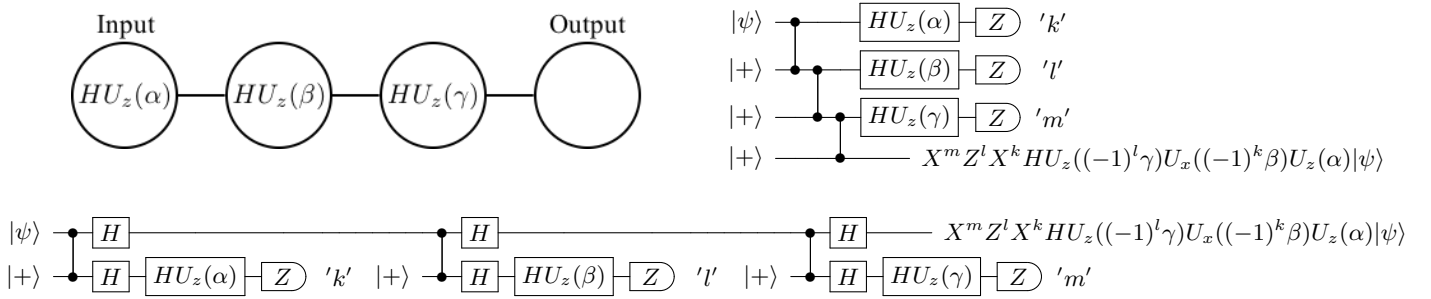
\begin{figure*}[t]
\setlength{\unitlength}{0.4cm}
\begin{picture}(5,15)

\put(-13,8){\includegraphics{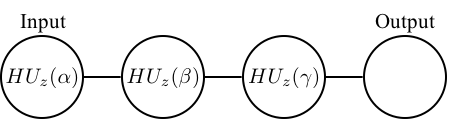}}
\end{picture}
\begin{picture}(5,0)
\put(3,12){\[
\Qcircuit @C=.7em @R=.4em {
\lstick{|\psi\rangle} & \ctrl{1} & \qw & \qw & \gate{HU_z(\alpha)} & \measureD{Z} & & 'k' \\
\lstick{|+\rangle} & \ctrl{-1} & \ctrl{1} & \qw & \gate{HU_z(\beta)} & \measureD{Z} & & 'l' \\
\lstick{|+\rangle} & \qw & \ctrl{-1} & \ctrl{2} & \gate{HU_z(\gamma)} & \measureD{Z} & & 'm' \\
\\
\lstick{|+\rangle} & \qw & \qw & \ctrl{-2} & \rstick{X^mZ^lX^kHU_z((-1)^l\gamma)U_x((-1)^k\beta)U_z(\alpha)|\psi\rangle} \qw
}
\]}
\end{picture}
\begin{picture}(6,5)
\put(-24,5){\[
\Qcircuit @C=.7em @R=.4em  {
\lstick{|\psi\rangle} & \ctrl{1} & \gate{H} & \qw & \qw & \qw & \qw & \qw & \qw & \qw & \qw & \qw & \ctrl{1} & \gate{H} & \qw & \qw & \qw & \qw & \qw & \qw & \qw & \qw & \qw & \ctrl{1} & \gate{H} & \rstick{X^mZ^lX^kHU_z((-1)^l\gamma)U_x((-1)^k\beta)U_z(\alpha)|\psi\rangle} \qw\\
\lstick{|+\rangle} & \ctrl{-1} & \gate{H} & \gate{HU_z(\alpha)} & \measureD{Z} & & 'k' & & & & & \lstick{|+\rangle} & \ctrl{-1} & \gate{H} & \gate{HU_z(\beta)} & \measureD{Z} & & 'l' & & & & & \lstick{|+\rangle} & \ctrl{-1} & \gate{H} & \gate{HU_z(\gamma)} & \measureD{Z} & & 'm'
}
\]}
\end{picture}
\vspace{-0.8cm}
\captionsetup{justification=RaggedRight,singlelinecheck=false}
\caption{An arbitrary single-qubit unitary computation over a linear cluster of input $|\psi\rangle$, propagated via Z-axis measurements (upper-left), with circuital representations of the standard (upper-right) and sequential (lower) protocols. The two circuits are equivalent when acting on pure states with ideal gates (see text for the definition of the notation). See Ref.~\cite{RAF+11} for the equivalence in case of a generic cluster with both finite- and infinite-dimensional systems.}
\label{StanSeqSchemata}
\end{figure*}

Despite the aforementioned advances, the cluster state itself presents a significant technical challenge: to carry out computation, one has to protect the entangled nodes from various sources of noise long enough not only to assemble the cluster state itself, but also to conduct all of the measurements required. In order to mitigate this issue, it is possible to exploit the fact that it is not necessary to assemble the entire cluster state at the beginning of the computation, but rather it can be continuously built ``on the fly'' \cite{LR09,MMR,menicucci,waddington,zwerger}. This can be viable in some implementations, in particular with travelling light \cite{furu,gershoni}, but cumbersome in other relevant contexts that involve stationary systems, such as trapped atoms, ions, or solid-state qubits.

More general approaches to computation similar in spirit to MBQC --- in the sense that they also rely on adaptable measurements to drive the computation --- have been considered in the literature, and are usually referred to as quantum computation via ancillae \cite{munro1, munro2, proctor1, proctor2, proctor3}. In particular, Anders \textit{et al.~}\cite{anders} have introduced a setting --- dubbed \textit{ancilla-driven quantum computation} (ADQC) --- in which the quantum systems involved in the computation are classified in two sets that mutually interact (see also Refs.~\cite{kashefi,morimae10}): the first one comprises \textit{registers} that are assumed to be kept in a protected environment allowing for long coherence times; the second one is composed of \textit{ancillae} that are continuously generated and consumed via measurements after having interacted with the registers. Similarly to MBQC, it is the pattern of such measurements that drives the computation and depends on the algorithm to be implemented, whereas, at difference with MBQC, the registers only store and update the computation results, without ever being measured. The resemblance between these two approaches can be actually formalised into a full equivalence if one restricts to the relevant case in which the ancillae interact with no more then two registers (and one time only per register). In fact, Roncaglia \textit{et al.} have shown that in this case ADQC is equivalent to the realisation of MBQC in which the cluster is generated on the fly and, in addition, the nodes are sequentially swapped at each interaction (via a computation independent local operation) \cite{RAF+11}. For this reason, we will refer to this specific model of ADQC as \textit{Sequential MBQC}. The given prescription is equivalent to the conventional MBQC model, hereafter referred to as \textit{Standard MBQC}, under ideal conditions (see Fig.~\ref{StanSeqSchemata}). Beyond this, little is known about the properties of sequential MBQC (and ADQC more in general) under the effects of noise and in realistic settings. The objectives of this work are first to analyse the impact of \textit{non}-ideal conditions on this equivalency, and second to determine the circumstances in which the standard or the sequential model is more resilient when subject to the detrimental action of given models of noise.

By construction, sequential MBQC is well suited for hybrid implementations in which two types of quantum systems, or generally two degrees of freedom, enjoy complementary characteristics: long coherence lifetime on one hand, ease of local manipulations and measurements on the other hand. Various experimental platforms could host such a hybrid scenario, with prominent examples given by cavity QED \cite{Haroche,ramanpulse}, where atoms hosted in cavities can play the role of registers, whereas injected light act as ancillae; or circuit QED \cite{cirQED}, with artificial atoms and microwave photons acting as registers and ancillae respectively. In the last part of this work, we will focus in detail on the first setting, highlighting the effect of experimentally relevant forms of noise in the implementation of sequential MBQC.

In section II, we provide a brief summary of sequential MBQC, and how it differs from its standard counterpart. In section III, we introduce sets of operations that may be concatenated for universal computation in both the standard and sequential formalisms, in addition to the noise models used for our analysis of non-ideal performance. This analysis is contained in section IV, with an examination of sequential performance in a cavity-QED setting in section V. Finally, we provide our conclusions in section VI.

The present work will deal only with the analysis of quantum  computation based on systems composed of qubits. We will use the following standard notation for qubit states, operations, and graphical representation of circuits:
\begin{figure}[!htb]
\centering
\begin{minipage}{.23\textwidth}
\[
\Qcircuit @C=.7em @R=.4em @! {
& & \gate{H} & \qw
}
\]
\end{minipage}%
\begin{minipage}{0.23\textwidth}
\vspace{.4cm}
Hadamard gate
\end{minipage}
\end{figure}

\begin{figure}[!htb]
\centering
\begin{minipage}{.23\textwidth}
\[
\Qcircuit @C=.7em @R=.4em @! {
& \gate{U_z(\alpha)} & \qw
}
\]
\end{minipage}%
\begin{minipage}{0.23\textwidth}
\vspace{.4cm}
Unitary rotation about the Z-axis of angle $\alpha$: $U_z(\alpha)=\exp[-\frac{i}{2} \alpha Z]$
\end{minipage}
\end{figure}

\begin{figure}[!htb]
\centering
\begin{minipage}{.23\textwidth}
\[
\Qcircuit @C=.7em @R=.4em @! {
& & & & \ctrl{1} & \qw \\
& & & & \\
& & & & \ctrl{-1} & \qw
}
\]
\end{minipage}%
\begin{minipage}{0.23\textwidth}
\vspace{.4cm}
Control-PHASE (CZ)
\end{minipage}
\end{figure}

\begin{figure}[!htb]
\centering
\begin{minipage}{.23\textwidth}
\vspace{-.5cm}
\[
\Qcircuit @C=.7em @R=.4em @! {
& \measureD{Z} \\
& \measureD{X}
}
\]
\end{minipage}%
\begin{minipage}{0.23\textwidth}
Measurements, conducted in the $\{|0\rangle\langle0|,|1\rangle\langle1|\}$ and $\{|+\rangle\langle+|,|-\rangle\langle-|\}$ basis respectively.
\end{minipage}
\end{figure}

\section{Sequential MBQC}

For standard MBQC, an entire \textit{cluster state} is generated, after which the computation is implemented by measuring qubits one after the other, where the basis of measurement selected in each case determines the precise operation to be carried out. In each case, these bases of measurement have to be corrected according to the fed-forward outcome of measurements immediately prior, establishing a causal order: consequently, this gives rise to the alternate title, the \textit{one-way model of computation} \cite{raussendorf}.

\begin{figure}[!h]
%\centering
\includegraphics[width=8cm]{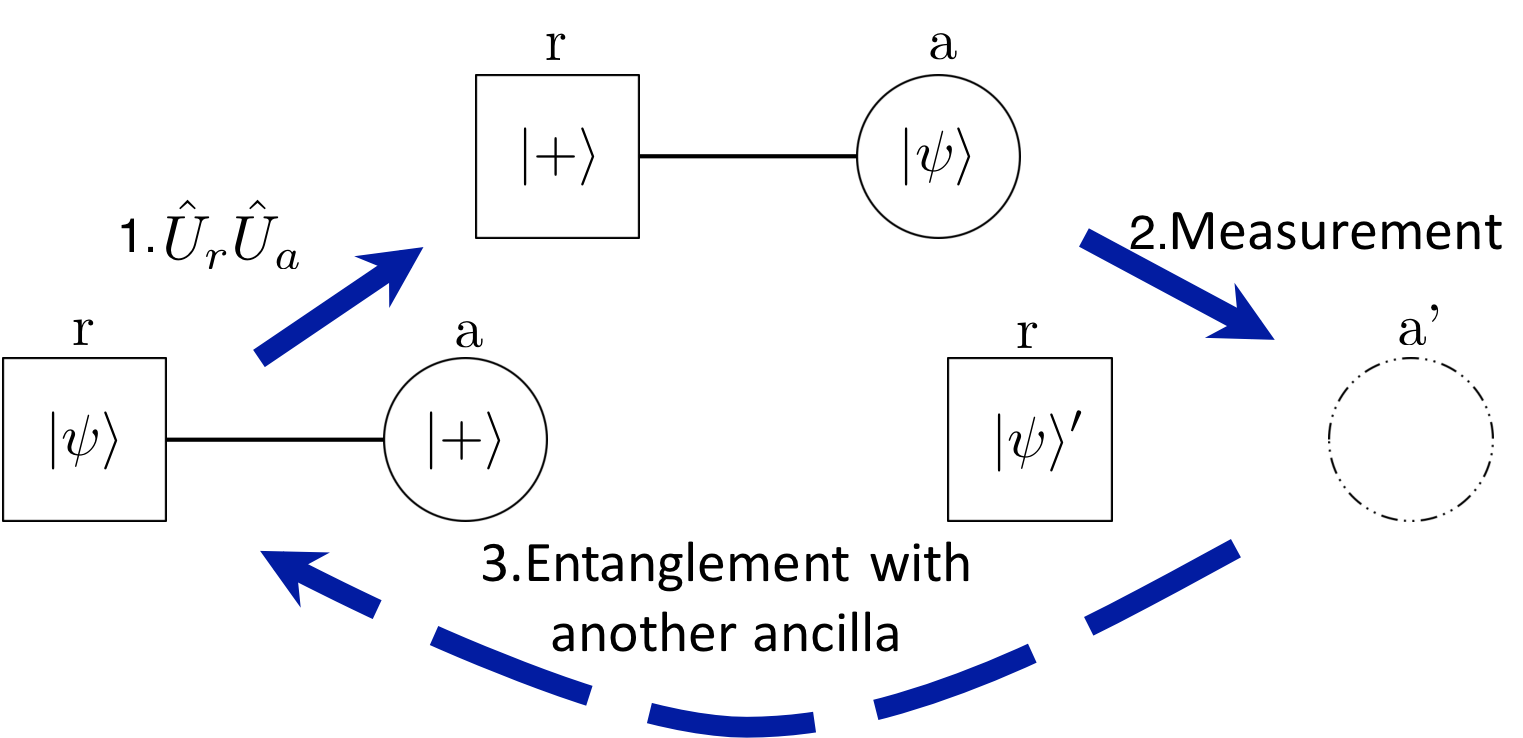}
\caption{The 3 steps of sequential MBQC}
\label{seqcycle}
\end{figure}

As said, in the sequential approach to MBQC, qubits are denoted as either flying "ancilla" nodes, or long-lived "register" nodes, the latter of which is host to the input state to be processed. The sequence of operations to be conducted is as follows (with pictorial form in Fig.\ref{seqcycle}):

\begin{enumerate}
\item An ancilla $a$, prepared in the superposition state $|+\rangle\equiv(|0\rangle+|1\rangle)/\sqrt{2}$, is entangled with the appropriate register node $r$ via a $CZ$ operation. 
\item The two entangled nodes undergo a unitary transformation $\hat{U}_T$ via \textit{local complementation} to swap their states. 
\item The ancilla is measured in the appropriate basis, teleporting the operated input state $|\psi\rangle'$ back onto the register.
\end{enumerate}

The net effect of this sequence is equivalent to the creation and consumption of a 2-qubit linear cluster in standard MBQC, with the exception that the freshly operated input state returns to its original node $r$ after measurement of the ancilla $a$. As a result, the sequence can be repeated using new ancillae until the desired output is created, without requiring extensive preservation of the ancillae or direct and adaptable operations on the registers.

Conveniently, the crux of sequential MBQC, the local complementation of entangled nodes, reduces to the local operation $H_r\otimes H_a$ between the participant nodes $r$ and $a$ for qubits \cite{RAF+11}. Furthermore, as the complementation of these nodes is typically immediately subsequent to their entanglement, the two operations can be combined into a bespoke "complemented-$CZ$" gate $\Gamma_{Seq}\equiv\left(H\otimes H\right)CZ$. This preserves the MBQC approach that computation is conducted via algorithm-independent entangling gates and measurements, with only the latter having to be adapted in order to drive the computation towards the desired outcome.

\section{Universal gates and noise models}
\label{tools}

Using the equivalency between circuit and cluster models of computation, it is possible to write circuit schematics for various measurement-based computations. To be specific, as the effects of measuring entangled nodes are equivalent to gate teleportation \cite{nielsenchuang}, key elements of MBQC can be depicted circuits implementing gate teleportations among up to 4 qubits each. To test and compare the two models of computations, we shall therefore consider the gate teleportation of all single-qubit Z-rotations and the $CX$ (control-NOT) operations. Regarding the former, the circuit representation of standard gate teleportation of $HU_z(\alpha)$ from the input node to an ancilla node is given by \cite{Nielsen06}
\[
\Qcircuit @C=.7em @R=.4em {
& & & & & & \lstick{|\psi\rangle} & \ctrl{1} & \gate{HU_z(\alpha)} & \measureD{Z} & & 'k' \\
& & & & & & \lstick{|+\rangle} & \ctrl{-1} & \qw & \rstick{X^kHU_z(\alpha)|\psi\rangle} \qw
}
\]

\noindent
where the by-product operation $X^k$ depends on the outcome of the measurement and it is applied in a successive step of the computation by adapting the corresponding measurement. Notice also that the Hadamard operations $H$ are ubiquitous in the standard MBQC approach. The circuit representation of sequential gate teleportation is instead given by
\[
\Qcircuit @C=.7em @R=.4em  {
& & & & \lstick{|\psi\rangle} & \ctrl{1} & \gate{H} & \qw & \rstick{X^kHU_z(\alpha)|\psi\rangle} \qw \\
& & & & \lstick{|+\rangle} & \ctrl{-1} & \gate{H} & \gate{HU_z(\alpha)} & \measureD{Z} & & 'k' 
}
\]

\noindent
where one can see that, unlike the standard version, the gate only is teleported onto the input system itself. One can see that the output states of both the sequential and standard schemes are equivalent in such an ideal setting.
 
Concerning the $CX$ gate, the circuit representation of its implementation via standard MBQC is given by \cite{raussendorf}
\[
\Qcircuit @C=.7em @R=.4em  {
& & & & & & & \lstick{|c\rangle} & \qw & \ctrl{2} & \qw & \qw & \qw & \rstick{Z^p|c\rangle} \\
& & & & & & & \lstick{|t\rangle} & \ctrl{1} & \qw & \qw & \measureD{X} & & 'p' \\
& & & & & & & \lstick{|+\rangle} & \ctrl{-1} & \ctrl{-2} & \ctrl{1} & \measureD{X} & & 'q' \\
& & & & & & & \lstick{|+\rangle} & \qw & \qw & \ctrl{-1} & \qw & \qw & \rstick{X^qZ^p|t\oplus c\rangle}
}
\]

\noindent
with the output state written in modulo 2. The sequential counterpart of this circuit is given by
\[
\Qcircuit @C=.7em @R=.4em  {
& \lstick{} & & & & & & & \hspace{0.4cm} |c\rangle & & \ctrl{3} & \gate{H} & \rstick{X^pH|c\rangle} \qw \\
& \lstick{|t\rangle} & \ctrl{1} & \gate{H} & \qw & \qw & \qw & \qw & \ctrl{2} & \gate{H} & \qw & \qw & \rstick{X^qZ^p|t\oplus c\rangle} \qw \\
& \lstick{|+\rangle} & \ctrl{-1} & \gate{H} & \measureD{X} & & 'p' \\
& \lstick{} & & & & & \hspace{-0.3cm} |+\rangle & & \ctrl{-2} & \gate{H} & \ctrl{-3} & \gate{H} & \measureD{Z} & & 'q'
}
\]
Notice that the original register nodes $|c\rangle$ and $|t\rangle$ never directly interact, yet they evolve as if fed into a $CX$ gate, up to local corrections $X$, $Z$ and $H$. Notice that again the output states of both the sequential and standard schemes are equivalent in such an ideal setting.

Together, these circuits implement gate teleportation of the set $\{U_z(\alpha), CX\}$ in both the standard and sequential models, which is made universal by recognising the indirect implementation of the $H$ gate by choosing to leave it uncorrected post-measurement. As any general computation can be implemented via concatenation of this set \cite{Nielsen06}, we expect any conclusions derived from the analysis of these circuits to be valid for the entirety of their respective models.

The detrimental effect of noise in the circuits above will be considered using various models. In particular we will use binary stochastic maps: either the procedure in question works as intended, or is disrupted by noise. As both the preparation and the entanglement of nodes are crucial to the running of cluster models, it is possible to witness significant perturbation by applying such a map to these two procedures.

For the preparation of ancillae, the noisy outcome in our model is the unintentional creation of the maximally mixed state:
\begin{equation}
\hspace{0.7cm}|+\rangle\langle+|\rightarrow(1-\eta)|+\rangle\langle+|+\eta\left(\frac{I}{2}\right),
\label{ancilla}
\end{equation}
where $I$ is the identity operator, and $\eta$ parametrises the noise in the ancilla's creation. 

Concerning the entangling gates $\Gamma$ (with $\Gamma_{sta}$ and $\Gamma_{seq}$ referring to the entangling gates in standard and sequential computation respectively), there are two different noisy models that we will consider: (i) either the two input systems are depolarized; or (ii) the gate simply misfires, leaving the input untouched. The corresponding maps acting on the generic 2-qubit state $|\phi\rangle$ are given by:
\begin{align}
\hspace{-0.5cm}
\Gamma|\phi\rangle\langle\phi|\Gamma^\dagger&\xrightarrow{\text{Depolarizing}}&(1-p)\Gamma|\phi\rangle\langle\phi|\Gamma^\dagger+p\left(\frac{I}{4}\right);
\label{CZdep}
\\
\hspace{-0.5cm}
\Gamma|\phi\rangle\langle\phi|\Gamma^\dagger&\xrightarrow{\text{\makebox[1.27cm]{Misfiring}}}&(1-s)\Gamma|\phi\rangle\langle\phi|\Gamma^\dagger+s|\phi\rangle\langle\phi|.
\label{CZmis}
\end{align}
where, similar to the ancilla creation above, the probability of depolarizing and misfiring is determined by $p$ and $s$ respectively.
Because different experimental incarnations of the entangling gates may be subject to one or the other of these potential models --- the \textit{depolarizing} and \textit{misfiring} models --- their effects will be analysed separately \cite{note_measure}.

In regards to quantifying the impact the above noise models have on computational output, we consider two figures of merit: the gate infidelity $G$ and the diamond distance. The former is a widely used measure of noise resilience \cite{nielsengate}:
\begin{align}
\hspace{0.6cm}& G(\mathcal{F},\mathcal{E}) \equiv1-F(\mathcal{F},\mathcal{E}) \nonumber
\\
\hspace{0.6cm}& =1-\left[\frac{1}{d+1}+\frac{\sum_j\text{Tr}[\mathcal{F}(U^\dagger_j)\mathcal{E}(U_j)]}{d^2(d+1)}\right]
\label{gateinfidelity}
\end{align}
where $d$ is the dimensionality of the system ($d=2$ for qubit operations), $U_j$ is an orthogonal operator basis in said system, and $\mathcal{F}$ and $\mathcal{E}$ are the maps corresponding to the action of ideal and noisy quantum circuits respectively. It has the major advantage of being a straightforward quantity to calculate in the case of qubit systems, often allowing one to obtain useful analytical expressions. Unfortunately it also hosts a set of relevant drawbacks \cite{sanders,flammia,wallman}. First, it does not take into account the effect that a noisy gate acting on a given qubit has on other qubits to which the former might be quantumly correlated --- a situation that is unavoidable in non-trivial computations. In addition, by taking the average over all possible inputs (accordingly to the Haar measure), the gate infidelity underestimates the negative effects that worst case scenarios have on a computation. Thirdly, it typically also underestimates the effect of non-stochastic errors in the implementation of quantum gates. For these reasons we will test our results also against the diamond distance, which remedies all the issues above, with the drawback that it is not amenable to analytical calculations. The diamond distance $D$ \cite{waltrous} is defined as:
\begin{equation}
\hspace{1.3cm}D(\Psi)\equiv||\Psi||_\diamond=\frac{1}{2}||\Psi\otimes\mathcal{I}||_1,
\label{diamondnorm}
\end{equation}
where $\Psi=\mathcal{F}-\mathcal{E}$, and the r.h.s. operator 1-norm of a map $\Phi$ is defined as $||\Phi||_1\equiv\text{max}\{||\Phi(X)||_1: X\in L(\mathcal{X}), ||X||_1\leq1\}$, where $L(\mathcal{X})$ is the set of linear operators within the complex Euclidean space $\mathcal{X}$, and $||X||_1=\text{Tr}\left(\sqrt{X^\dagger X}\right)$ is the trace norm of an operator. For the difference between two maps, this metric can be physically interpreted as a geometric distance between the two computational processes represented by said maps. A useful property of the diamond distance is its tensoring with an auxiliary Hilbert space, allowing it to implicitly account for noise that can be modelled via coupling to an external system, such as information leakage to an external environment. In addition, the diamond distance innately accounts for \textit{worst-case scenario} noise, as opposed to the average-case noise computed by various (in)fidelity measures \cite{flammia}. As the fault-tolerance of a computational process is directly linked to its worst-case performance, it logically follows that a metric such as the diamond distance is more appropriate for the comparison of potentially fault-tolerant models of computation.

\section{Comparison between standard and sequential MBQC}

We will now compare the performances of standard and sequential models in the setting given above --- namely, when they are used to implement a set of universal operations in a noisy environment. We will distinguish two major scenarios. First, we will consider perfect Hadamard operations throughout both models, henceforth referred to as the \textit{perfect complementation} scenario. This is motivated by the fact that (i) Hadamard gates are ubiquitous in both cases, including in measurements and by-product operations, and (ii) the major sources of noise in experiments are typically due to imperfect state preparation and entangling gates. Second, we will consider the \textit{imperfect complementation} scenario, featuring imperfect Hadamard operations for the sequential model only, whenever they act as companions of the $CZ$ operations --- namely, when they implement local complementation as part of $\Gamma_{Seq}$. This can be thought of as the worst case scenario for the sequential model, and it is motivated by the fact that the entangling gate $\Gamma_{Sta}$ used in the standard model is just given by the $CZ$ operation ($\Gamma_{Sta}=CZ$) whereas the entangling gate in the sequential model is always, as said above, complemented by two Hadamard gates: $\Gamma_{Seq}=(H\otimes H)CZ$. Therefore, for some implementations, it might be more convenient to directly build $\Gamma_{Seq}$ than to complement a $CZ$ with two local Hadamard gates. In this case, noise acts on $\Gamma_{Seq}$ itself, in turn implying that both $\Gamma_{Sta}$ and $\Gamma_{Seq}$ have to be treated on equal footing.

\subsection{Perfect local complementation}

We start our comparison between the two models using the gate infidelity as figure of merit. We have computed the latter for the full noise range $0<\{\eta,p,s\}<1$ for both models across the minimal set of operations required for universality --- analysing relative resilience is then simply a matter of taking the difference between the standard and sequential gate infidelities. As an analytical metric, we were able to derive exact expressions for the four circuits of section \ref{tools}. Subtracting the standard infidelity $G_{\rm stan}$ from the sequential one $G_{\rm seq}$ for the four cases studied above, we respectively obtain:

\begin{center}
\begin{tabular}{|p{2cm}|p{6.5cm}|}
\hline
& $G_{\rm seq}-G_{\rm stan}$ (Perfect Complementation)\\
\hline
$U_z(\alpha)$ & $\frac{1}{3}\eta(p-1)\cos^2{\alpha}\label{GFPa}$ \\
Depolarizing & \\
\hline
$U_z(\alpha)$ & $\frac{1}{6}\left[\eta s-\eta-s+\eta(s-1)\cos{2\alpha}\right]\label{GFPb}$ \\
Misfiring & \\
\hline
$CX$ & $\frac{1}{20}(p-1)^2\left\lbrace4\eta(\eta-2)\right. $ \\ Depolarizing & $\left. +p\left[1-4\eta(\eta-2)\right]\right\rbrace\label{GFPc}$ \\
\hline
$CX$ & $\frac{1}{10}\left\lbrace s \left[s(6-s)-2 \right]\right. $ \\ Misfiring & $\left. -2\eta\left[2-s(6-s(7-2s)) \right]-2\eta^2(s-1)^3 \right\rbrace\label{GFPd}$ \\
\hline
\end{tabular}
\end{center}

Note that the equations for $U_z(\alpha)$ are strictly non-positive, regardless of rotation angle $\alpha$: consequently, sequential 1-qubit gate teleportation \textit{always} yields superior resilience (smaller gate infidelity), regardless of the type of noise considered.

\begin{figure}[!b]
    \centering
    \begin{minipage}{.23\textwidth}
    	\textbf{(a)}
	\includegraphics[width=\linewidth]{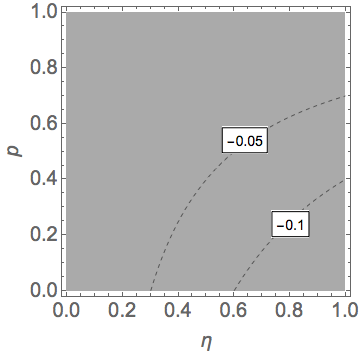}
	\label{OneDepoGF_LCP}
    \end{minipage}
    \begin{minipage}{0.23\textwidth}
    	\textbf{(b)}
	\includegraphics[width=\linewidth]{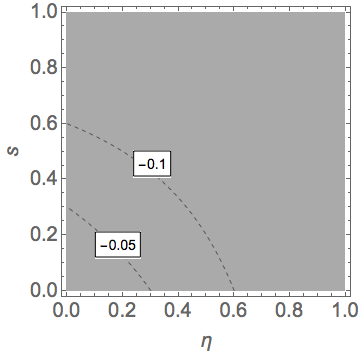}
	\label{OneMisGF_LCP}
    \end{minipage}
    \begin{minipage}{.23\textwidth}
    	\textbf{(c)}
	\includegraphics[width=\linewidth]{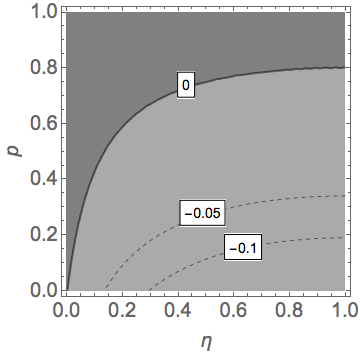}
	\label{TwoDepoGF_LCP}
    \end{minipage}
    \begin{minipage}{0.23\textwidth}
    	\textbf{(d)}
	\includegraphics[width=\linewidth]{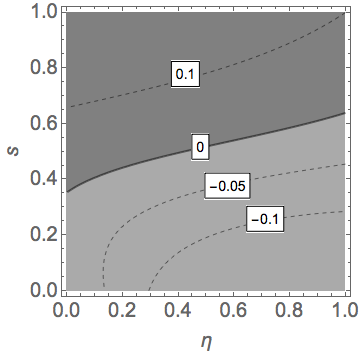}
	\label{TwoMisGF_LCP}
    \end{minipage}
    	 \captionsetup{justification=RaggedRight,singlelinecheck=false}
   	 \caption{Difference in standard/sequential gate infidelities, with perfect sequential local complementation, for: (a) Teleporting T gate, with depolarizing noise; (b) Teleporting T gate, with misfiring noise; (c) Teleporting $CX$, with depolarizing noise; (d) Teleporting $CX$, with misfiring noise.}
    	\label{FidelityLCP}
\end{figure}

For the case of $T$ and $CX$ gates [$T\equiv U_z(\pi/4)$], these differences have also been plotted in Fig. \ref{FidelityLCP}, where darker regions indicate noise ranges that the standard model is more resilient to noise, whereas negative differences are a lighter shade, and correspond to areas the sequential approach is more robust.

\begin{figure}[!b]
    \centering
    \begin{minipage}{.23\textwidth}
    	\textbf{(a)}
	\includegraphics[width=\linewidth]{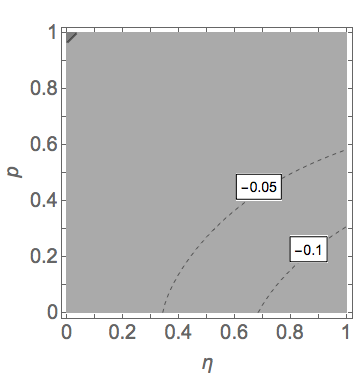}
	\label{OneDepoDD_LCP}
    \end{minipage}
    \begin{minipage}{0.23\textwidth}
    	\textbf{(b)}
	\includegraphics[width=\linewidth]{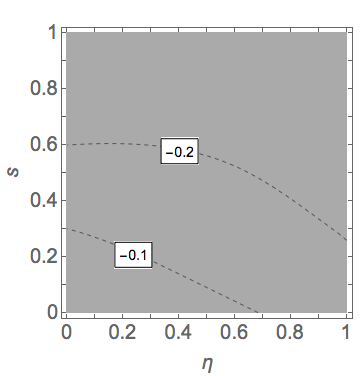}
	\label{OneMisDD_LCP}
    \end{minipage}
    \begin{minipage}{.23\textwidth}
    	\textbf{(c)}
	\includegraphics[width=\linewidth]{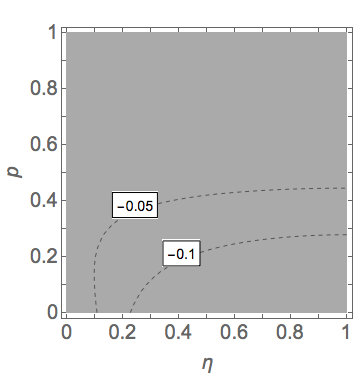}
	\label{TwoDepoDD_LCP}
    \end{minipage}
    \begin{minipage}{0.23\textwidth}
    	\textbf{(d)}
	\includegraphics[width=\linewidth]{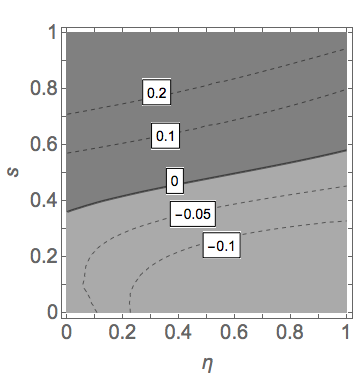}
	\label{TwoMisDD_LCP}
    \end{minipage}
    \captionsetup{justification=RaggedRight,singlelinecheck=false}
    \caption{Difference in standard/sequential diamond distances, with perfect sequential local complementation, for: (a) Teleporting T gate, with depolarizing noise; (b) Teleporting T gate, with misfiring noise; (c) Teleporting $CX$, with depolarizing noise; (d) Teleporting $CX$, with misfiring noise.}
    \label{DiamondLCP}
\end{figure}

Figs.~\ref{FidelityLCP}(c) and \ref{FidelityLCP}(d) show that, unlike the $T$ gate, the infidelity for the $CX$ gate exhibits shifting bias towards the standard models as levels of $\Gamma_{seq}$ noise increase (\textit{i.e.}, as $p$ or $s$ increase). However, the standard model appears to exhibit superior resilience only for noise values where the gate performance is already compromised and therefore not of practical interest. In particular, regarding the misfiring noise, when $s<0.35$ the sequential model is always superior [subplot \ref{FidelityLCP}(d)]; for depolarizing noise on the other hand, this is true whenever $p<\frac{4(2\eta-\eta^2)}{1+8\eta-4\eta^2}$ [subplot \ref{FidelityLCP}(c), zero level curve].

The prior results are given using the gate infidelity as figure of merit. However, as said, a more appropriate choice is represented by the diamond distance, for which the results are given in Fig.~\ref{DiamondLCP}. Generally, we can see that these confirm the ones obtained using the gate infidelity, meaning that the latter is a rather good proxy for the scenarios under consideration. However there are some appreciable differences between Figs.~\ref{FidelityLCP} and \ref{DiamondLCP}, the most relevant of which concerns the implementation of the $CX$ gate under depolarizing noise, namely subplot (c). As said above, the gate infidelity individuates only one relevant region of parameters for which the standard model is apparently more resilient then the sequential one. From the subplot in question, one can see that this is no longer the case when the diamond distance is considered: in fact, the two subplots disagree on whether standard MBQC can exhibit superior performance \textit{at all}. In other words, while the gate infidelity is a far easier measure to compute, there are configurations that exhibit major deviations from results yielded by the diamond distance, compromising its potential as an approximate projection of performance. Summarizing, the diamond-distance based assessment reveals that the sequential approach is more resilient to noise for all the parameters of relevance for quantum computation (namely, the low noise regions in all the subplots of Fig.~\ref{DiamondLCP}). 

\subsection{Imperfect local complementation}

As said, while the entanglement operation $CZ$ has to be implemented identically regardless of the model, local complementation in sequential MBQC exists as a distinct operation with no counterpart in the standard model. Consequently, the treatment of this operation given above (namely, the absence of noise for $H$) can be unmotivated in some implementations, thus introducing an unfair advantage to the sequential model with respect to the standard one. If local complementation is non-ideal, then the sequential model is host to a type of noise non-existent in standard MBQC; if local complementation is ideal, then approximately half of all sequential operations are immediately guaranteed to succeed. The latter scenario has been considered in the previous section: we now analyze the former, which is in this sense the worst-case scenario for sequential MBQC. We will therefore explore now the relative performance between standard and sequential MBQC if depolarizing and misfiring noise were to affect not only the $CZ$ gate, but the $\Gamma_{sta}$ and $\Gamma_{seq}$ gates in the standard and sequential approach respectively.

\begin{figure}[!t]
    \centering
    \begin{minipage}{.23\textwidth}
    	\textbf{(a)}
	\includegraphics[width=1\linewidth]{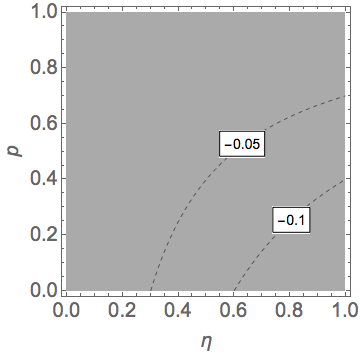}
	\label{OneDepoGF}
    \end{minipage}
    \begin{minipage}{0.23\textwidth}
    	\textbf{(b)}
	\includegraphics[width=1\linewidth]{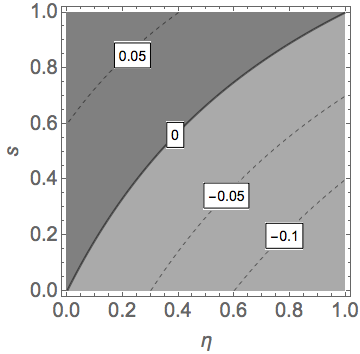}
	\label{OneMisGF}
    \end{minipage}
    \begin{minipage}{.23\textwidth}
    	\textbf{(c)}
	\includegraphics[width=1\linewidth]{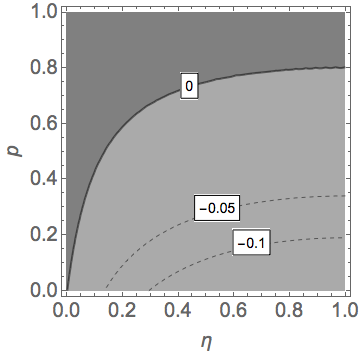}
	\label{TwoDepoGF}
    \end{minipage}
    \begin{minipage}{0.23\textwidth}
    	\textbf{(d)}
	\includegraphics[width=1\linewidth]{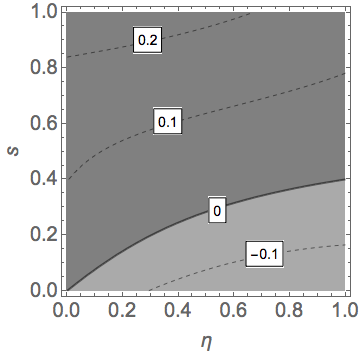}
	\label{TwoMisGF}
    \end{minipage}
    	 \captionsetup{justification=RaggedRight,singlelinecheck=false}
    	\caption{Difference in standard/sequential gate infidelities, with imperfect sequential local complementation, for: (a) Teleporting T gate, with depolarizing noise; (b) Teleporting T gate, with misfiring noise; (c) Teleporting $CX$, with depolarizing noise; (d) Teleporting $CX$, with misfiring noise.}
    	\label{fidelities}
\end{figure}

\begin{figure}[!t]
    \centering
    \begin{minipage}{.23\textwidth}
	\textbf{(a)}
	\includegraphics[width=1\linewidth]{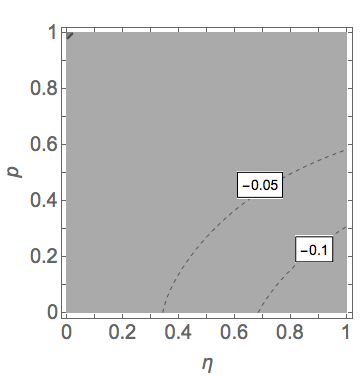}
	\label{OneDepoContrast}
    \end{minipage}
    \begin{minipage}{0.23\textwidth}
    	\textbf{(b)}
	\includegraphics[width=1\linewidth]{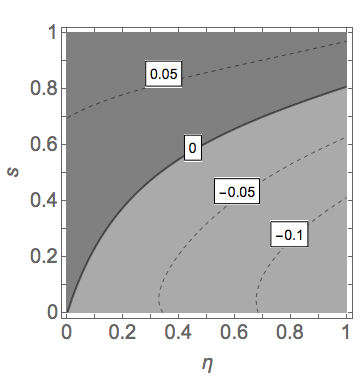}
	\label{OneMisContrast}
    \end{minipage}
    \begin{minipage}{.23\textwidth}
    	\textbf{(c)}
	\includegraphics[width=1\linewidth]{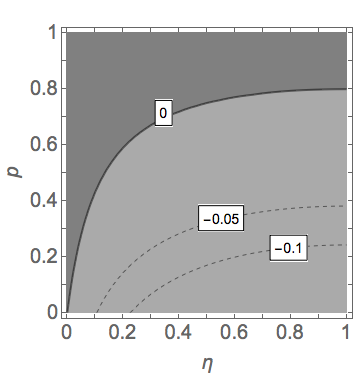}
	\label{TwoDepoContrast}
    \end{minipage}
    \begin{minipage}{0.23\textwidth}
    	\textbf{(d)}
	\includegraphics[width=1\linewidth]{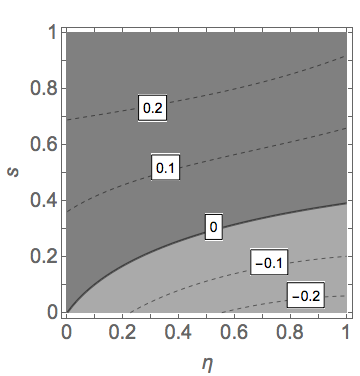}
	\label{TwoMisContrast}
    \end{minipage}
    	 \captionsetup{justification=RaggedRight,singlelinecheck=false}
    	\caption{Difference in standard/sequential diamond distances, with imperfect sequential local complementation, for: (a) Teleporting T gate, with depolarizing noise; (b) Teleporting T gate, with misfiring noise; (c) Teleporting $CX$, with depolarizing noise; (d) Teleporting $CX$, with misfiring noise.}
    	\label{distances}
\end{figure}

Considering the gate infidelity as measure of noise, we are able as before to derive exact expressions for the four circuits, with $\eta$, $p$, and $s$ representing ancilla, depolarizing, and misfiring noise as before. Subtracting the standard infidelity $G_{\rm stan}$ from the sequential one $G_{\rm seq}$ for the four cases studied above, we respectively obtain:
\begin{center}
\begin{tabular}{|p{2cm}|p{6.5cm}|}
\hline
& $G_{\rm seq}-G_{\rm stan}$ (Imperfect Complementation)\\
\hline
$U_z(\alpha)$ & $\frac{1}{3}\eta(p-1)\cos^2{\alpha}\label{GFa}$ \\
Depolarizing & \\
\hline
$U_z(\alpha)$ & $\frac{1}{6}\left[s+\eta(s-2)\right]\cos^2{\alpha}\label{GFb}$ \\
Misfiring & \\
\hline
$CX$ & $\frac{1}{20}(p-1)^2\left\lbrace 4\eta(\eta-2)\right. $ \\ Depolarizing & $+\left. p\left[1-4\eta(\eta-2)\right] \right\rbrace\label{GFc}$ \\
\hline
$CX$ & $\frac{1}{20}\left\lbrace 2s \left[ 5+s(7s-9) \right]- \eta(9s^3-14s+8)\right. $ \\ Misfiring & $\left. +2\eta^2(s-1)(3s-2) \right\rbrace\label{GFd}$ \\
\hline
\end{tabular}
\end{center}

Regarding direct observations from the equations themselves, sequential teleportation of $U_z(\alpha)$ now only hosts superior performances for \textit{all} parameters \textit{only} if the $CZ$ gates are host to depolarizing noise.

As before, we give the contour plots corresponding to the above results for the $T$ and $CX$ gates in Fig.\ref{fidelities}. We can see that the regions of noise for which the sequential approach yields superior performance have decreased in both size and magnitude: this is to be expected, given that the noise in $\Gamma_{seq}$ now affects all the operations included in it (\textit{i.e.}, both $CZ$ and $H$).

The impact of including imperfect local complementation in the depolarizing and misfiring noise models can be clearly observed comparing Figs. \ref{FidelityLCP} and \ref{fidelities}. The most evident change in behaviour is for the teleportation of the $T$ gate under misfiring noise [subplots (b)]: sequential bias is now maximal in the low-$s$ / high-$\eta$ range of noise, as opposed to high-$s$ / high-$\eta$ as in the perfect local complementation scenario.

The comparison using the diamond distance is given in Fig.~\ref{distances}, and mainly agrees with the findings obtained using the gate infidelity --- confirming that the latter is a relatively good proxy of the former in this setting.

For the most part, sequential MBQC hosts a smaller diamond distance for the majority of potential noise values. The exception is for $CX$ teleportation over a misfiring $CZ$ channel, where the standard model is superior whenever entangling noise is non-negligible. Consequently, any quantum computers utilising MBQC that may exhibit misfiring $CZ$ gates must take the quality of their ancillae and entangling procedures into consideration when determining whether to utilise standard or sequential protocols.

Summarizing the results above for both perfect and imperfect local complementation, one can conclude that the sequential model is more resilient to noise than the standard one for low rates of entangling failure, a configuration relevant to fault-tolerant computation. Physically this can be intuitively understood by considering that, whereas in the sequential approach quantum information is stored always in the same physical systems, in the standard approach it is continuously transferred from one system to another. Namely, at every step of the computation the information is physically teleported from one set of nodes of the cluster state to the next ones. On the other hand, it is true that also in the sequential approach a swap of quantum information is needed at every step. What the analysis above proves quantitatively is that, in the interesting regimes, the latter operation introduces less noise with respect to the one central to the standard approach.

(For reference, we have included the gate infidelities \& diamond distances of the sequential circuits used to generate the results of this section in the appendix.)

\section{Experimental feasibility in a cavity-QED setting}

Having shown in the previous Section that the sequential approach is often more resilient to noise than the standard one, it is now relevant to consider specific physical implementations of the former. In particular, we will focus on cavity-QED settings, which represent a natural choice for its experimental realisation.

To elaborate, recall that the sequential model designates qubits as either longer-lived registers or expendable ancillae: ideal candidates for these categories in the context of cavity QED are atoms trapped in cavities and circularly-polarized photons respectively. As optical operations and measurements can be performed with an exceptional degree of accuracy (\textit{e.g.}, see Refs. \cite{LHF_Bell} where photonic systems are used to demonstrate significant loop-hole free tests of Bell's theorem), we will focus in more details on the noise that affects the atomic gates.

In short, atoms are manipulated via Raman pulses \cite{ramanpulse} which rotate their state about one of the Bloch sphere's main axes by an angle determined by the duration of the pulse. General rotations are made possible by using Raman pulses in sequence --- for our purposes, a rotation about the Z axis by $\pi$ followed by a Y-rotation of $\pi/2$ is equivalent to a Hadamard operation up to an overall phase.
%\begin{align}
%R_y(\pi/2)R_z(\pi) & =\left(\begin{array}{cc}1/\sqrt{2} & -1/\sqrt{2} \\ 1/\sqrt{2} & 1/\sqrt{2}\end{array}\right)\left(\begin{array}{cc}-i & 0 \\ 0 & i\end{array}\right) \nonumber
%\\ & =-i\frac{1}{\sqrt{2}}\left(\begin{array}{cc}1 & 1 \\ 1 & -1\end{array}\right)
%\label{ramanH}
%\end{align}

Consequently, a mistiming in the duration for which a Raman pulse acts on the cavity translates into an offset to the angle of its rotation. Instead of a stochastic map such as the ones used for flawed ancilla preparation and $CZ$ operations, we will simulate the effect of these offset angles by selecting them from a normal distribution, using its standard deviation $\gamma$ as a proxy for the quality of control over the pulse duration. This noise model is better-suited for representing systematic errors in the calibration of the Raman pulse apparatus.
\begin{figure}[!t]
    \centering
    \begin{minipage}{.4\textwidth}
    	\textbf{(a)}
	\includegraphics[width=1\linewidth]{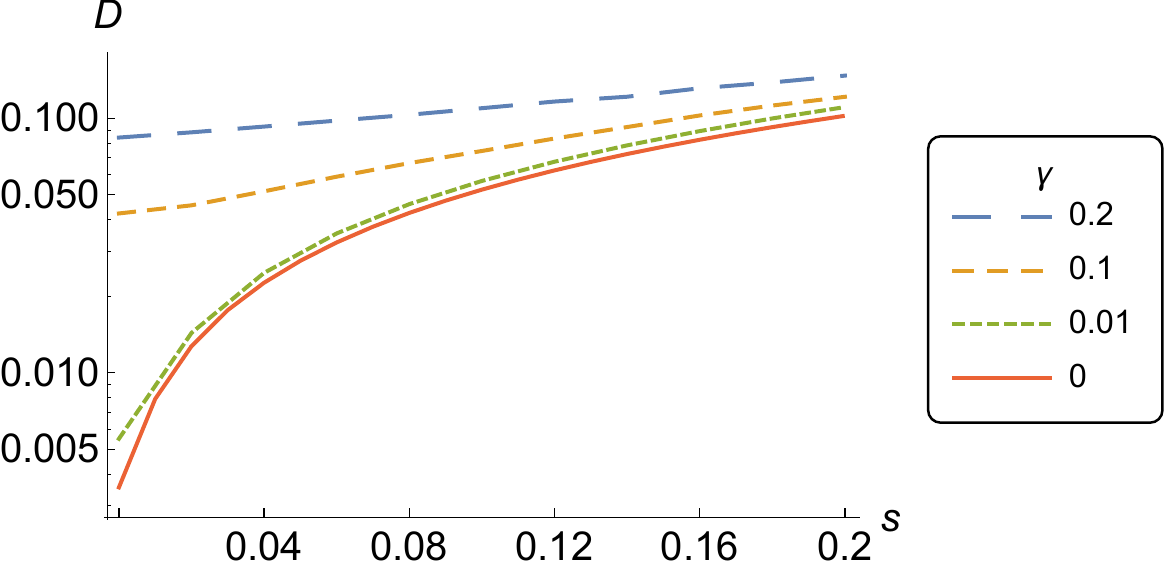}
	\label{OneExpGF}
    \end{minipage}
    \begin{minipage}{0.4\textwidth}
    	\textbf{(b)}
	\includegraphics[width=1\linewidth]{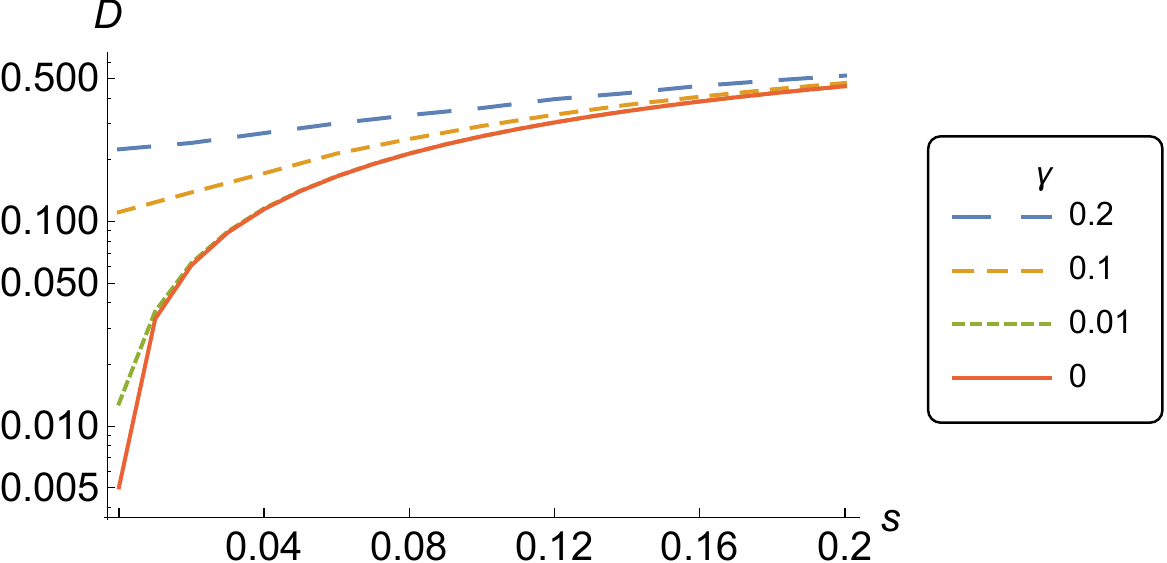}
	\label{TwoExpGF}
    \end{minipage}
    \begin{minipage}{0.4\textwidth}
    	\textbf{(c)}
	\includegraphics[width=1\linewidth]{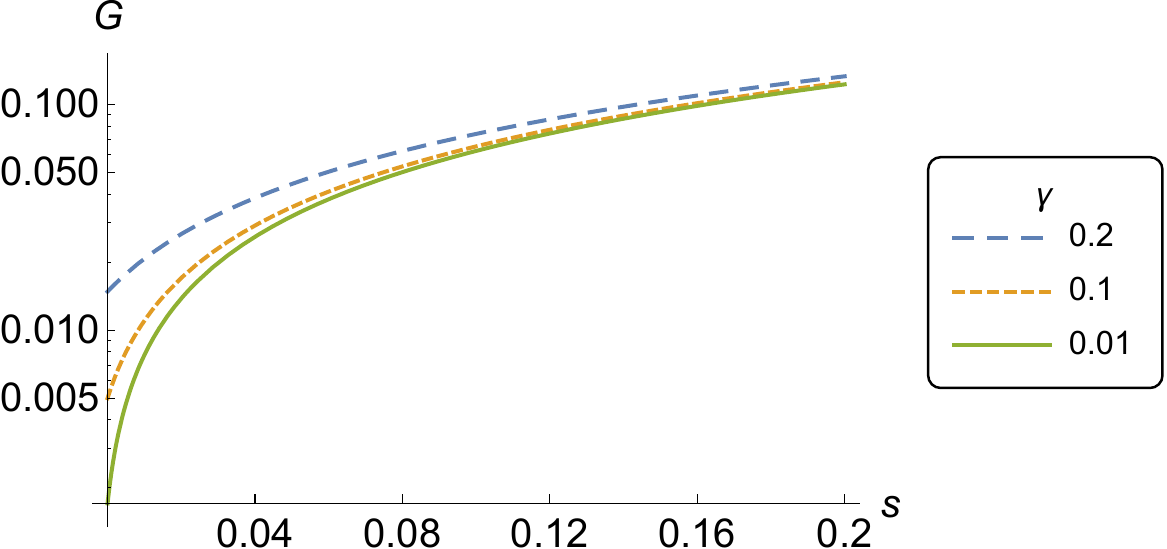}
	\label{OneExpDD}
    \end{minipage}
    \begin{minipage}{0.4\textwidth}
    	\textbf{(d)}
	\includegraphics[width=1\linewidth]{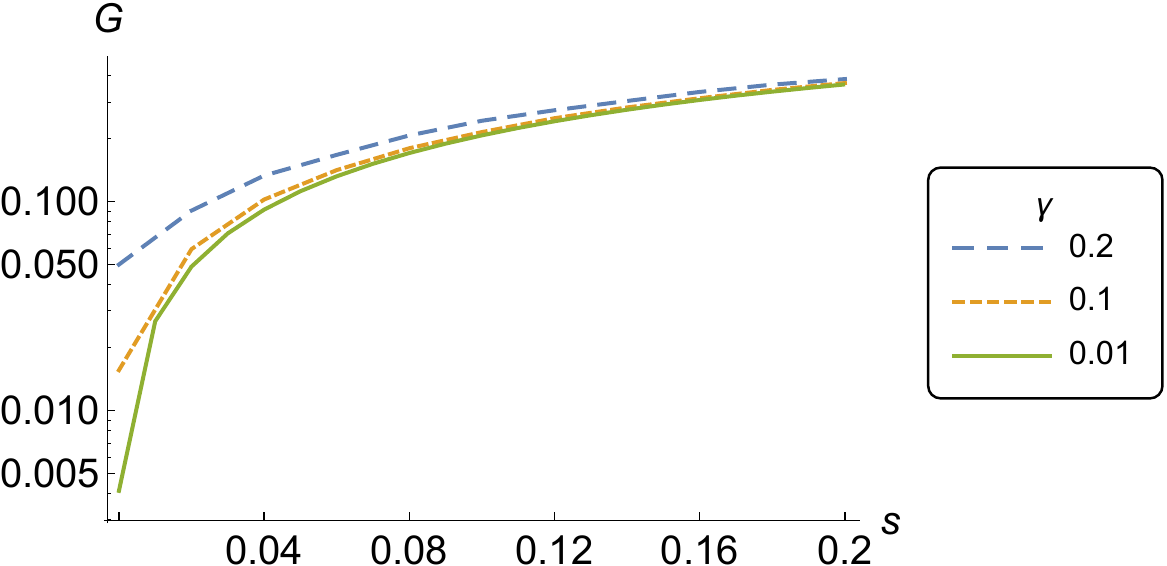}
	\label{TwoExpDD}
    \end{minipage}
   \captionsetup{justification=RaggedRight,singlelinecheck=false}
        \caption{Diamond distances $D$ (a,b) and gate infidelities $G$ (c,d) of the experimental simulation of the Gate Teleportation of (a,c) $T$; (b,d) $CX$ gates. Vertical axes use logarithmic scaling. Misfiring noise is restricted to $s\leq0.2$, ancilla noise fixed as $\eta=0.01$.}
        \label{experimental}
\end{figure}

In addition, in consideration of feasibility, constraints are applied to the types of noise previously considered. First, we fix the ancilla noise $\eta$ to 1\%, compatible with the high level of purity achievable in photonic experiments. Second, we will consider only misfiring noise for the CZ operations, since this is the noise model that better describes what is observed in cavity-QED experiments \cite{reiserempe}. The reason for this is that a major systematic error that arises in photon-cavity interactions is the reflection of the photon from the cavity, without interacting with the contained atom --- that is to say, the intended interaction "misfires" (conversely, the cavity/photon interaction is not affected by a relevant amount of depolarizing noise). We will set an upper limit of $s=0.2$ to the misfiring noise, which corresponds to the maximum fidelity of 0.83 reported for experimental entanglement of similar systems, with some additional room for error \cite{reiserempe}.

The results for both diamond distance $D$ and gate infidelity $G$ are reported in Fig.~\ref{experimental}. The latter can be derived analytically from Eq.~(\ref{gateinfidelity}) and, for the case of $T$ gate teleportation, it shows a linear dependence on the noise parameter $s$:
\begin{equation}
G=\frac{1}{48}e^{-\gamma^2}\left(a_0+a_1 s\right)\;,
\label{GFE}
\end{equation}
with
\begin{align}
a_0 &=
4 \left[2 + 6 e^{\gamma^2 }- e^{\gamma^2/2} (\eta-4) - \eta\right] \;,\\
a_1 &= 4 e^{\gamma^2/2} (\eta-4) + 4 (\eta-2) - e^{\gamma^2} (5 + \eta) \;.
\label{GFE_aux}
\end{align}

Qualitatively, it is evident that the flawed Raman pulse has a detrimental impact on both the diamond distance and gate infidelity. Quantitatively, there is a marked deviation between the two measures as $s$ approaches zero: looking at the results for $\gamma=0.2$ in Fig.~\ref{experimental} (the blue [upper dashed] line), the reported diamond distance and gate infidelity for $s=0$ differ by a factor of approximately 4. It is clear that the systematic nature of potential Raman pulse offsets can produce extremal scenarios that the gate infidelity fails to fully encapsulate, compromising its validity for quantitative analysis of such noise. This is in line with what was observed by Kueng \textit{et al.} \cite{flammia}, where it is noted that, for small experimental errors in detuning and calibration, the worst-case behavior can be orders of magnitude worse than the average-case one. As said, the former is captured by $D$ and the latter by $G$, thus explaining the marked difference between diamond distance and gate infidelity observed in Fig.~\ref{experimental} for small $s$. 

These results clearly demonstrate that, despite the sequential approach to MBQC being generally more resilient to noise with respect to the standard approach, this scheme still requires working regimes that are demanding if compared to the ones that are experimentally achievable with state-of-the-art technology. 

\section{Conclusions}

Our analysis demonstrates that sequential MBQC exhibits superior resilience to noise, compared to the standard approach to MBQC for a large range of scenarios: assuming the local complementation operation of sequential MBQC can be implemented independently of faulty entangling gates, this statement is general for all the regimes of interest for quantum computation (namely, low levels of noise). Nonetheless, by comparing with state-of-the-art cavity-QED settings, we have also shown that the requirements for achieving high-quality gates via this approach are demanding. Our results can in fact be used as benchmarks for future experiments.

Furthermore, we have assessed gate performances by using both the gate (in)fidelity and the diamond distance. We have shown that the more widespread measure of the gate (in)fidelity only yields results coincident with the diamond distance on a qualitative scale, making necessary to evaluate the latter in scenarios assessing gate performances against fault-tolerant thresholds. In particular, we have seen that the possibility for systematic sources of noise requires the use of the diamond distance to fully quantify the extent to which computational performance may be disrupted.

A general conclusion that is evident from the analysis reported here is that the equivalence between sequential and standard MBQC breaks down in non-ideal settings. This means that when it comes to perform a fault-tolerant analysis of sequential MBQC, one cannot simply apply the results obtained for the standard model. In particular, given that the sequential model typically exhibits a superior resilience, we expect that implementations based on it could meet more easily the stringent requirements demanded by fault-tolerant quantum computation. This, along with the effects of imperfect measurements \cite{morimae10}, will be topics of further investigations that we plan to perform in the future. Also, we plan to conduct similar investigations using continuous-variable systems as opposed to qubits \cite{RAF+11,proctor3}, to clarify whether the choice of the dimensionality of the physical system supporting the computation has an impact on the relative resilience of computational models.

\section{Acknowledgements}

The authors acknowledge funding from the UK-EPSRC (Award Reference	1494765).

\newpage
\appendix

\section{Computing the Diamond Distance}

A major downside to the diamond distance is that there is no obvious means of computing it directly for a given difference in maps $\Delta$ --- consequently, it must be obtained numerically. Thankfully, paradigms such as semi-definite optimisation \cite{flammia} exist that can carry out this task efficiently.

Some initial preparation is required. The map $\Delta$ must first be converted into a superoperator matrix $J(\Delta)$ via the \textit{Choi-Jamiolkowski representation} \cite{waltrous}:
\begin{equation}
J(\Delta)=\sum_{0\leq j,k\leq 1}\Delta(|j\rangle\langle k|)\otimes|j\rangle\langle k|,
\label{choimatrix}
\end{equation}
where the set of $\{|j\rangle\}$ give a basis for the Hilbert space under consideration. Using this Choi matrix $J(\Delta)$, the diamond distance of the original map $\Delta$ can be computed using an appropriately-written Semi-Definite Program (SDP), such as the one presented below \cite{flammia}:
\begin{tabular}{lllll}
\\
\textbf{Primal problem}
\\
\\Maximize: & $\langle J(\Delta),W\rangle$
\\Subject to: & $W\leq \rho\otimes I_d$,
\\ & $\text{Tr}(\rho)=1$,
\\ & $W\in\text{Pos}(A\otimes B)$,
\\ & $\rho\in\text{Pos}(A)$.
\\
\\\textbf{Dual problem}
\\
\\Minimize: & $||\text{Tr}_B(Z)||_\mathcal{1}$
\\Subject to: & $Z\geq J(\Delta)$,
\\ & $Z\in\text{Pos}(A\otimes B)$.
\label{semidefprogram}
\end{tabular}

In this prescription, $\langle X,Y\rangle=\text{Tr}(X^\dagger Y)$ is the Hilbert-Schmidt inner product of the matrices X and Y, $\text{Pos}(A\otimes B)$ denotes the cone of positive semidefinite operators acting on the space $A\otimes B$, and $||X||_\infty$ is equal to the largest eigenvalue of positive semidefinite X (also known as the operator norm of X). The optimal value for both the Primal and Dual problems, $\lambda$, is equal to $\frac{1}{2}||\Delta||_\Diamond$.

\section{Sequential gate infidelities \& diamond distances}

Overleaf we present for reference the gate infidelities and diamond distances we obtained for sequential gate teleportation of T and $CX$, for both depolarizing and misfiring entanglement, and both perfect and imperfect local complementation. Lighter regions correspond to lower infidelity/distance, and thus greater performance.

\begin{figure}%[!h]
    \centering
    \begin{minipage}{.23\textwidth}
	\textbf{(a)}
	\includegraphics[width=1\linewidth]{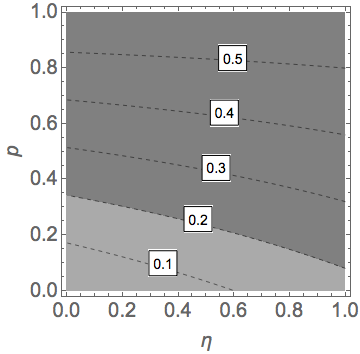}
	\label{OneDepoSeq_GF_LCP}
    \end{minipage}
    \begin{minipage}{0.23\textwidth}
    	\textbf{(b)}
	\includegraphics[width=1\linewidth]{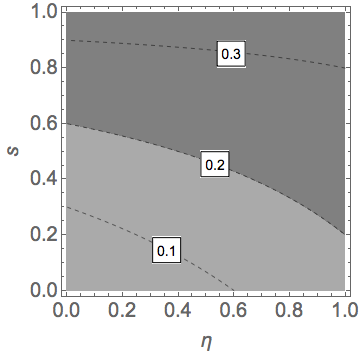}
	\label{OneMisSeq_GF_LCP}
    \end{minipage}
    \begin{minipage}{.23\textwidth}
    	\textbf{(c)}
	\includegraphics[width=1\linewidth]{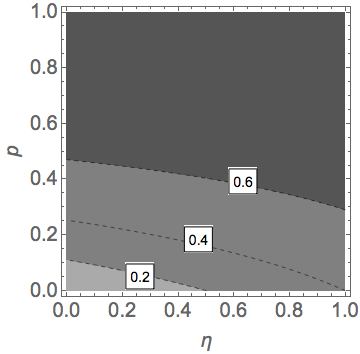}
	\label{TwoDepoSeq_GF_LCP}
    \end{minipage}
    \begin{minipage}{0.23\textwidth}
    	\textbf{(d)}
	\includegraphics[width=1\linewidth]{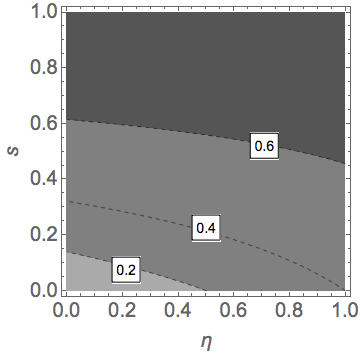}
	\label{TwoMisSeq_GF_LCP}
    \end{minipage}
    	 \captionsetup{justification=RaggedRight,singlelinecheck=false}
    	\caption{Sequential gate infidelities, with perfect local complementation, for: (a) Teleporting T gate, with depolarizing noise; (b) Teleporting T gate, with misfiring noise; (c) Teleporting $CX$, with depolarizing noise; (d) Teleporting $CX$, with misfiring noise.}
    	\label{seq_GF_LCP}
\end{figure}

\begin{figure}[h]
    \centering
    \begin{minipage}{.23\textwidth}
	\textbf{(a)}
	\includegraphics[width=1\linewidth]{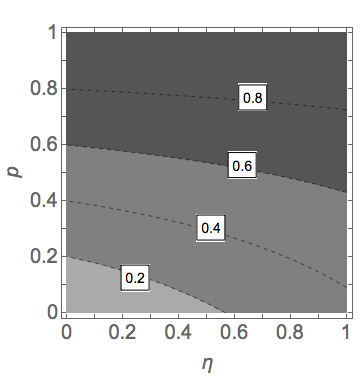}
	\label{OneDepoSeq_DD_LCP}
    \end{minipage}
    \begin{minipage}{0.23\textwidth}
    	\textbf{(b)}
	\includegraphics[width=1\linewidth]{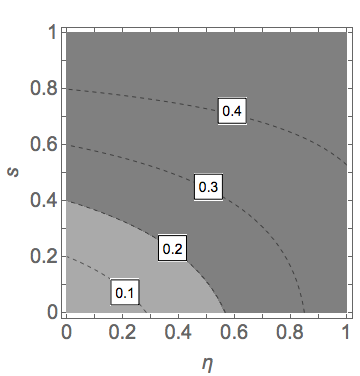}
	\label{OneMisSeq_DD_LCP}
    \end{minipage}
    \begin{minipage}{.23\textwidth}
    	\textbf{(c)}
	\includegraphics[width=1\linewidth]{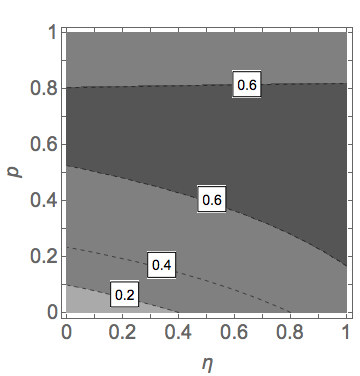}
	\label{TwoDepoSeq_DD_LCP}
    \end{minipage}
    \begin{minipage}{0.23\textwidth}
    	\textbf{(d)}
	\includegraphics[width=1\linewidth]{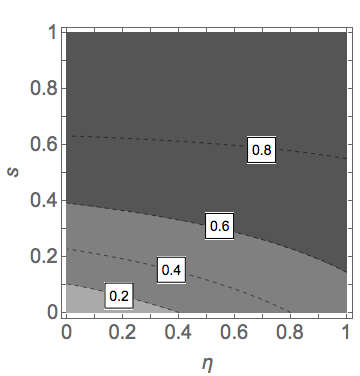}
	\label{TwoMisSeq_DD_LCP}
    \end{minipage}
    	 \captionsetup{justification=RaggedRight,singlelinecheck=false}
    	\caption{Sequential diamond distances, with perfect local complementation, for: (a) Teleporting T gate, with depolarizing noise; (b) Teleporting T gate, with misfiring noise; (c) Teleporting $CX$, with depolarizing noise; (d) Teleporting $CX$, with misfiring noise.}
    	\label{seq_DD_LCP}
\end{figure}

\begin{figure}%[!t]
    \centering
    \begin{minipage}{.23\textwidth}
	\textbf{(a)}
	\includegraphics[width=1\linewidth]{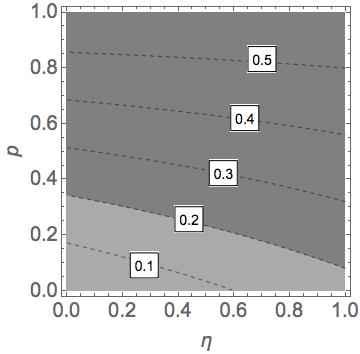}
	\label{OneDepoSeq_GF}
    \end{minipage}
    \begin{minipage}{0.23\textwidth}
    	\textbf{(b)}
	\includegraphics[width=1\linewidth]{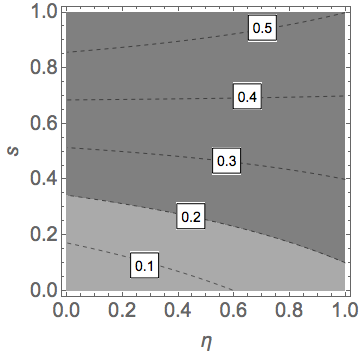}
	\label{OneMisSeq_GF}
    \end{minipage}
    \begin{minipage}{.23\textwidth}
    	\textbf{(c)}
	\includegraphics[width=1\linewidth]{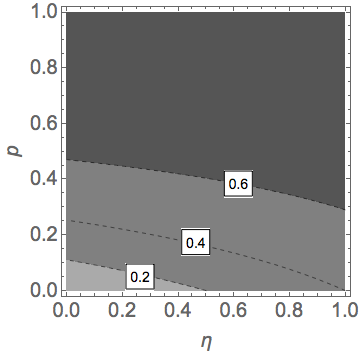}
	\label{TwoDepoSeq_GF}
    \end{minipage}
    \begin{minipage}{0.23\textwidth}
    	\textbf{(d)}
	\includegraphics[width=1\linewidth]{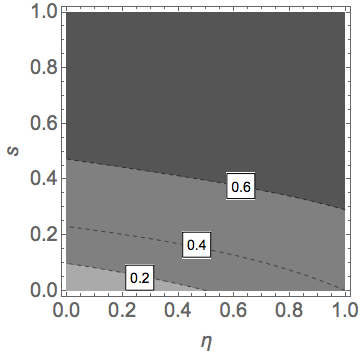}
	\label{TwoMisSeq_GF}
    \end{minipage}
    	 \captionsetup{justification=RaggedRight,singlelinecheck=false}
    	\caption{Sequential gate infidelities, with imperfect local complementation, for: (a) Teleporting T gate, with depolarizing noise; (b) Teleporting T gate, with misfiring noise; (c) Teleporting $CX$, with depolarizing noise; (d) Teleporting $CX$, with misfiring noise.}
    	\label{seq_GF}
\end{figure}

\begin{figure}%[!t]
    \centering
    \begin{minipage}{.23\textwidth}
	\textbf{(a)}
	\includegraphics[width=1\linewidth]{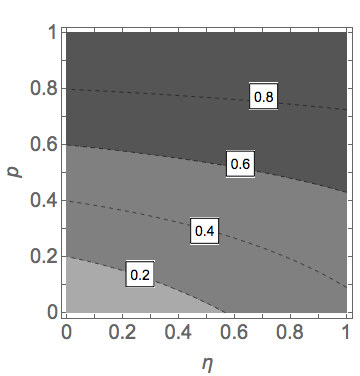}
	\label{OneDepoSeq_DD}
    \end{minipage}
    \begin{minipage}{0.23\textwidth}
    	\textbf{(b)}
	\includegraphics[width=1\linewidth]{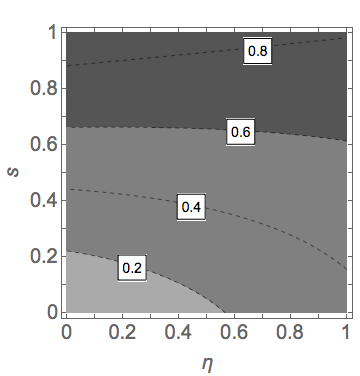}
	\label{OneMisSeq_DD}
    \end{minipage}
    \begin{minipage}{.23\textwidth}
    	\textbf{(c)}
	\includegraphics[width=1\linewidth]{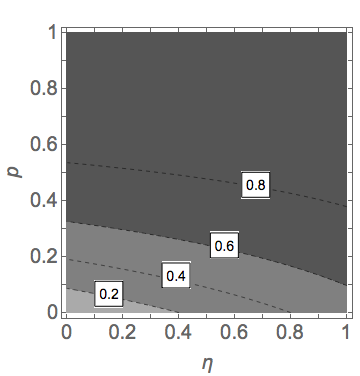}
	\label{TwoDepoSeq_DD}
    \end{minipage}
    \begin{minipage}{0.23\textwidth}
    	\textbf{(d)}
	\includegraphics[width=1\linewidth]{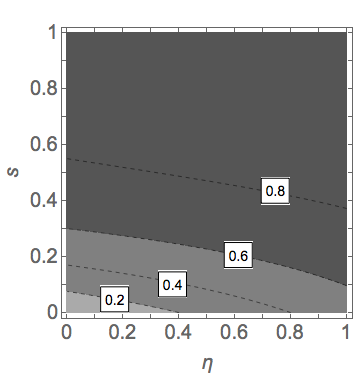}
	\label{TwoMisSeq_DD}
    \end{minipage}
    	 \captionsetup{justification=RaggedRight,singlelinecheck=false}
    	\caption{Sequential diamond distances, with imperfect local complementation, for: (a) Teleporting T gate, with depolarizing noise; (b) Teleporting T gate, with misfiring noise; (c) Teleporting $CX$, with depolarizing noise; (d) Teleporting $CX$, with misfiring noise.}
    	\label{seq_DD}
\end{figure}


\begin{thebibliography}{}
\bibitem{note_measure} It is quite possible for the last key ingredient of MBQC --- the measurements themselves --- to be affected by noise: the basis for a measurement may be perturbed about the Bloch sphere, for example. However, as the act of measurement itself is identical in both the standard and sequential models, it is assumed for convenience that measurement is an ideal, noise-free procedure.
\end{thebibliography}
\end{document}